\documentclass[prl,twocolumn]{revtex4}
\usepackage{graphicx}
\usepackage{amssymb,amsmath}
\usepackage{enumerate}

\newcommand{\be}{\begin{eqnarray}}
\newcommand{\ee}{\end{eqnarray}}

\begin{document}

\title{Distributed chaos and solitons at the edges of magnetically confined
plasmas }

\author{A. Bershadskii}

\affiliation{
ICAR, P.O. Box 31155, Jerusalem 91000, Israel
}

\begin{abstract}

It is shown, using results of measurements of ion saturation current in the plasma edges of different magnetic fusion confinement devices (tokamaks and stellarators), that the plasma dynamics in the edges is dominated by distributed chaos with spontaneously broken translational symmetry at low magnetic field, and with spontaneously broken reflexional symmetry (by helical solitons) at high magnetic field.

\end{abstract}

\maketitle

\section{Introduction}

Turbulent motion at the edges of magnetically confined plasmas in the magnetic fusion confinement devices degrades their performance. It is believed that space localized coherent structures (blobs) are the main cause for the confinement degradation. Direct measurements support the idea that these blobs are vortex-like velocity patterns associated with current filaments (with life span $\sim$ microseconds) \cite{spo}.

 The power spectra of ion saturation current is an effective tool in order to obtain information 
about the dynamic processes at the edges. Fortunately these spectra have a similar 
functional shape in most toroidal fusion devices (see, for instance, Ref. \cite{ped} where a survey of the data from several stellarators and JET tokamak are presented). This observation indicates that 
there could be a common underlying physical process. Presumably (see Ref. \cite{mm1} 
for a comprehensive analysis) this dynamics is chaotic rather than stochastic. 
The conclusion is based on the exponential-like shape of the observed spectra. Indeed, 
the exponential power spectra are observed in many dynamical systems with chaotic dynamics \cite{fm}-\cite{sig}. The exponential spectra can be provided \cite{mm2} by pulses having a Lorentzian functional form
$$
L(t)= \frac{A}{2} \left[\frac{\tau}{\tau + i(t-t_0)} + \frac{\tau}{\tau - i(t-t_0)}\right]  \eqno{(1)}
$$ 
For a series of $N$ Lorentzian pulses the power spectrum can be presented as a sum over the residues of the complex time poles
$$
E(\omega) \propto \left|\sum_{n=1}^{N}\exp (i\omega t_{0,n} - \omega \tau_n )\right|^2  \eqno{(2)}
$$
An exponential spectrum results from Eq. (2) if distribution of the Lorentzian pulse widths 
$\tau$ is sufficiently narrow one
$$
E(\omega) \propto \exp -(2\omega \tau)             \eqno{(3)}
$$
In the case of a broad (continuous) distribution of the pulse widths: $P(\tau )$, a weighted superposition of the exponentials Eq. (3) can approximate the sum Eq. (2) 
$$
E(\omega ) \propto \int_0^{\infty} P(\tau) \exp -(2\omega \tau) d\tau    \eqno{(4)}
$$

\section{Spectrum of distributed chaos}

  It is suggested in Ref. \cite{b1} that in the distributed chaos the weighted superposition Eq (4) is converged to a {\it stretched} exponential. Replacing $\tau$ by a dimensionless variable: $s=\tau/\tau_{\beta}$ ($\tau_{\beta}$ is a constant):
$$
E(\omega ) \propto \int_0^{\infty} P(s) \exp -s(2\omega \tau_{\beta})~  ds    \eqno{(5)}
$$ 
we can write the stretched exponential as
$$
E(\omega ) \propto \exp -(\omega/\omega_{\beta})^{\beta} \eqno{(6)}
$$
where $\omega_{\beta} = (2 \tau_{\beta})^{-1}$.

In the case of $\beta=1/2$ (this case is of a special interest for us, see below)
$$
P(s) \propto \frac{1}{s^{3/2}} \exp-(c/s) \eqno{(7)}
$$
where $c$ is a constant.

 For other values of $\beta$ the distribution $P(s,\beta)$ has rather cumbersome form \cite{jon}.

  In the experiments time signals are usually taken by probes at a fixed spatial location and reflect characteristics of the spatial structures moving past the probes. Therefore, the frequency spectra observed in the plasma experiments reflect the wavenumber spectra (so-called Taylor hypothesis \cite{my}) and the frequency spectrum Eq. (6) corresponds to wavenumber spectrum
$$
E(k ) \propto \exp -(k/k_{\beta})^{\beta} \eqno{(8)}
$$

\section{Asymptotic consideration}

\begin{figure}
\begin{center}
\includegraphics[width=8cm \vspace{-0.8cm}]{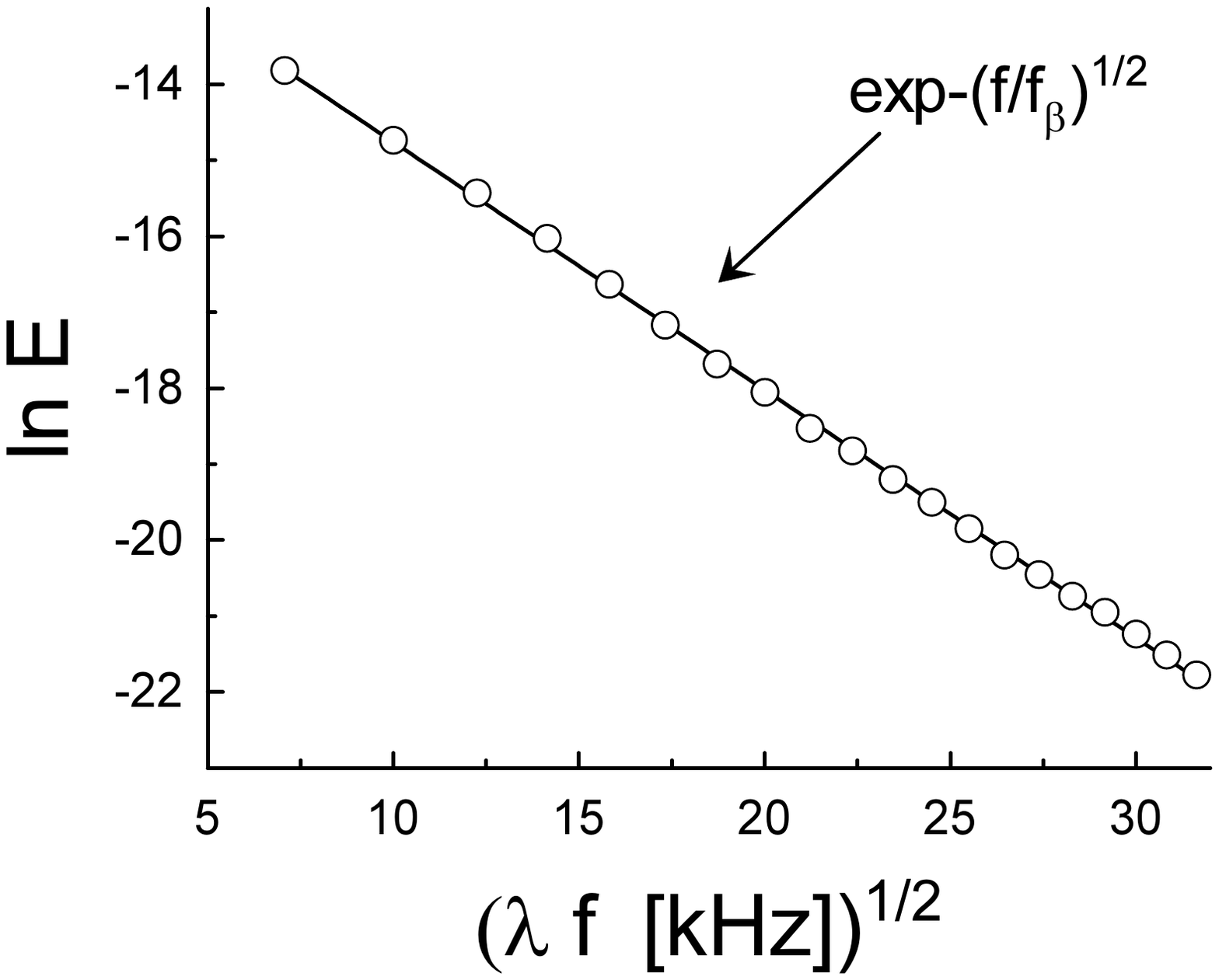}\vspace{-4.5cm}
\caption{\label{fig1} Power spectrum of ion saturation current in the plasma edge of different magnetic fusion confinement devices (tokamaks and stellarators). The frequency $f$ was rescaled by the authors of the Ref. \cite{ped} with the parameter $\lambda$: TJ-I: $\lambda =1$, TJ-IU: $\lambda =3$, JET: $\lambda =4.5$,  $W7-AS^{(1)}:~~\lambda = 3.5$, $W7-AS^{(2)}:~~\lambda = 4.5$. The straight line is drawn in order to indicate a stretched exponential decay Eq. (6) with $\beta =1/2$.} 
\end{center}
\end{figure}

The distribution $P(s,\beta )$ has rather simple asymptotic at $s \rightarrow 0$ \cite{jon}
$$
P(s,\beta ) \propto \frac{1}{s^{1+\beta/2(1-\beta)}} \exp-\frac{c}{s^{\beta/(1-\beta)}} \eqno({9)}
$$
Let the group velocity $\upsilon (\kappa )$ of the waves driving the distributed chaos be scale invariant at $\kappa \rightarrow \infty $ 
$$
\upsilon (\kappa ) \propto \kappa^{\alpha}     \eqno{(10)}
$$
and let $ \upsilon (\kappa ) $ to have a Gaussian distribution at $\kappa \rightarrow \infty $ : $~ \propto \exp-(\upsilon (\kappa )/\upsilon_b)^2$. In this case the asymptotic of $\kappa$ distribution is
$$
P(\kappa ) \propto \kappa^{\alpha -1}~\exp-\left(\frac{\kappa}{\kappa_b}\right)^{2\alpha}  \eqno{(11)}
$$ 
where $k_b$ is a constant.

Since $s \propto \kappa^{-1}$ then substituting Eqs. (9) and (11) into equation 
$$
P(s)~ds \propto P(\kappa)~ d\kappa    \eqno{(12)}
$$
one obtains
$$
\beta = \frac{2\alpha}{1+2\alpha}  \eqno{(13)}
$$

\section{Spontaneous breaking of space translational symmetry}

There exist two main space symmetries: rotational (isotropy) and translational (homogeneity). The 
Noether's theorem relates the conservation laws of the angular and linear momentum to these symmetries (respectively) \cite{ll2}. These conservation laws produce two hydrodynamic invariants: Loitsyanskii ($I_4$) and Birkhoff-Saffman ($I_2$) integrals \cite{my},\cite{bir}-\cite{dav2}:
$$   
I_n = \int r^{n-2} \langle {\bf u} ({\bf x},t) \cdot  {\bf u} ({\bf x} + {\bf r},t) \rangle d{\bf r}  \eqno{(14)}
$$ 
at sufficiently rapid decay of the two-point correlation function of the fluid velocity field. In the distributed chaos these two invariants drive two attractors. The attractors have different basins of attraction. The basin of attraction of the Loitsyanskii attractor is small and thin in comparison with that of the Birkhoff-Saffman attractor. Therefore, the scaling Eq. (10) is determined by the Birkhof-Saffman invariant ($I_2$) for the statistically stationary isotropic homogeneous turbulence  \cite{b1}:
$$
\upsilon (\kappa ) \propto ~I_2^{1/2}~\kappa^{3/2} \eqno{(15)}
$$
Then Eq. (13) provides $\beta =3/4$ \cite{b1}. 

\begin{figure}
\begin{center}
\includegraphics[width=8cm \vspace{-1.4cm}]{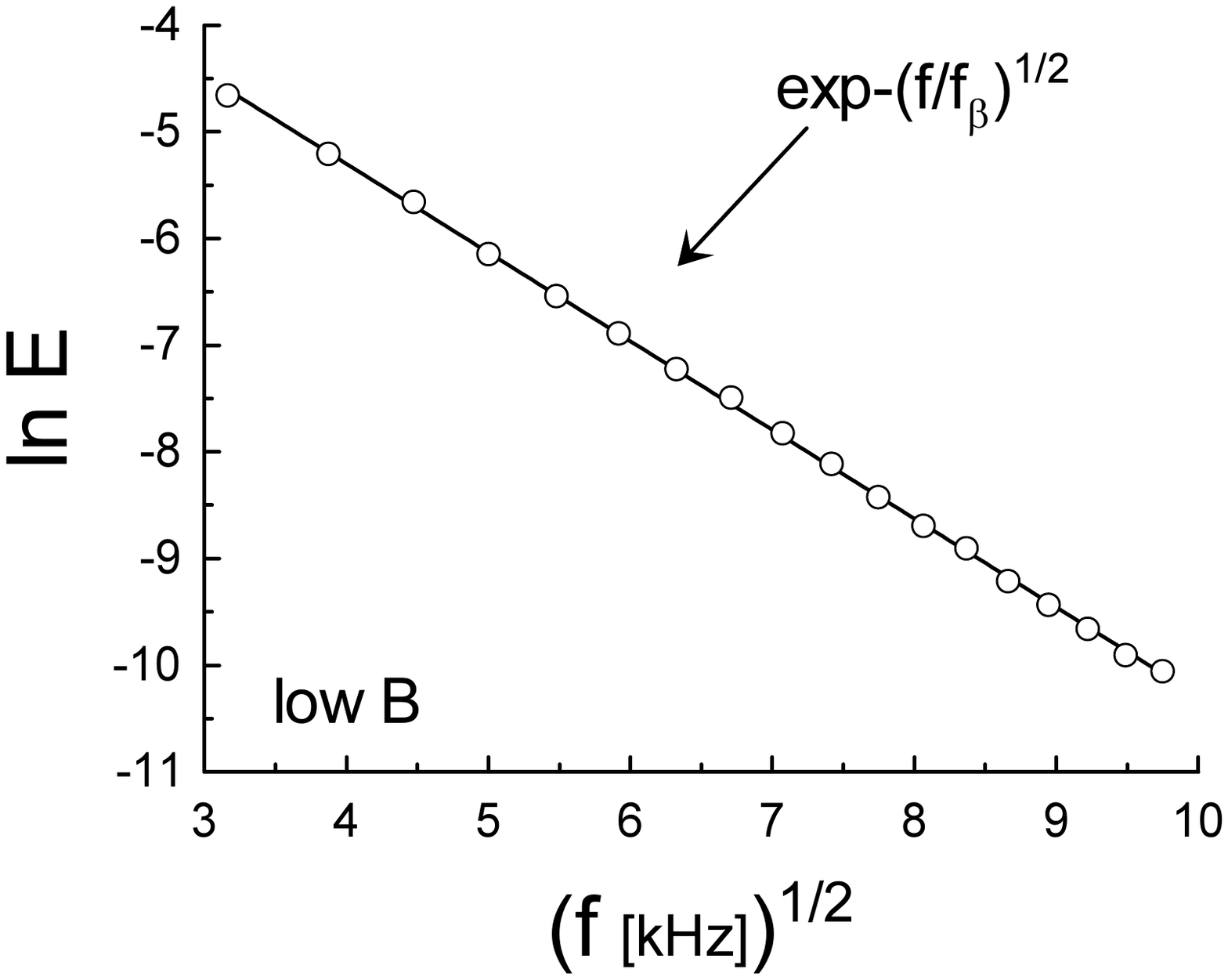}\vspace{-3.3cm}
\caption{\label{fig2} The same as in Fig. 1 but for TJ-K stellarator (a toroidal device). The data were taken from Ref. \cite{horn} and corresponds to the low magnetic field case (the edge region of the core plasma).} 
\end{center}
\end{figure}

  It is shown in Ref. \cite{b2} that at spontaneous breaking of the space translational symmetry 
the Birkhof-Saffman integral $I_2$ should be replaced in the scaling relationship Eq. (15) by a new integral 
$$
\gamma = \int_{V} \langle {\boldsymbol \omega} ({\bf x},t) \cdot  {\boldsymbol \omega} ({\bf x} + {\bf r},t) \rangle_{V}  d{\bf r} \eqno{(16)}
$$
where index $V$ means integration and averaging over the volume of motion and ${\boldsymbol \omega} ({\bf x},t)= \nabla \times {\bf u}  ({\bf x},t)$ is vorticity. Then the scaling relation Eq. (15) should be replaced by 
$$
\upsilon (\kappa ) \propto |\gamma|^{1/2}~\kappa^{1/2} \eqno{(17)}
$$
In this case Eq. (13) provides $\beta =1/2$ \cite{b2}.

Recent generalization of the Birkhoff-Saffman and Loitsyanskii invariants for magneto-hydrodynamic (MHD) turbulence \cite{dav1},\cite{dav2} allows to generalize the above consideration for MHD turbulence as well. 

  Figure 1 shows frequency power spectra of ion saturation current measured in the plasma edge of different fusion devices (tokamaks and stellarators). The frequency is rescaled by the authors of the Ref. \cite{ped} (by the parameter $\lambda$, see the legend to the figure) and the amplitudes are normalized in order to converge the data for the different devices in a single (universal) curve. These (cumulative) data for the Fig. 1 were taken from the Ref. \cite{mm1}, where the data were represented in a log-linear form. The scales in Fig. 1 are chosen in order to represent the Eq. (6) with $\beta =1/2$ as a straight line. The $\lambda$-rescaling of frequency $f$ takes into account that the parameter $f_{\beta}$ (or $\omega_{\beta}=2\pi f_{\beta}$) can be different for different devices. 
  
  Figure 2 shows results of more recent measurements performed in the edge region of the
core plasma at the TJ-K stellarator (a toroidal device) \cite{horn}. In this experiment low magnetic field corresponds to 72 mT and high magnetic field corresponds to 244 mT. Fig. 2 shows data obtained at the low magnetic field. As in Fig. 1 the scales in Fig. 2 are chosen in order to represent the Eq. (6) with $\beta =1/2$ as a straight line.
 
\section{Soliton driven distributed chaos} 

In strong magnetic field the waves driving the distributed chaos can be dominated by helical solitons (see, for instance, Ref. \cite{b3}). It is suggested in \cite{b3} that the spontaneously emerging pairs of helical solitons having opposite helical charges (spontaneous breaking of reflexional symmetry) can serve as 'sinks' of energy. In this case, unlike the Kolmogorov's-like turbulence, the energy dissipation rate should be taken not per volume but per soliton. And, correspondingly, dimension of the parameter governing this situation - $G$, should be $m^5/s^3$ \cite{b3}.  In the statistically stationary isotropic homogeneous turbulence the Birkhoff-Saffman integral {\it directly} determines the low wavenumbers power spectrum of velocity field \cite{saf}. Analogously in the soliton dominated MHD turbulence under strong magnetic field the low wavenumbers power spectrum of the velocity field is directly determined by the soliton parameter $G$ \cite{b3}:
$$
E(k) \propto G^{2/3} k^{1/3}   \eqno{(18)}
$$
 
Figure 3 (adapted from \cite{b3}) shows a streamwise power spectrum of the velocity in turbulent flow of mercury past a grid (bars of the grid were inclined in direction of the strong magnetic field). In the log-log scales the scaling law Eq. (18) corresponds to the straight line in the Fig. 3. The data were taken from an experiment reported in Ref. \cite{ps}. This result was  confirmed by analogous experiment reported in more recent Ref. \cite{bra}.  
 
 As for the waves driving the distributed chaos at the helical solitons domination 

$$
\upsilon (\kappa ) \propto ~G^{1/3}~\kappa^{2/3} \eqno{(19)}
$$ 
Then Eq. (13) provides $\beta =4/7$. 

 Figure 4 shows results of the same measurements as in Fig. 2 but for the high magnetic field.  The scales in Fig. 4 are chosen in order to represent the Eq. (6) with $\beta =4/7$ as a straight line.

\begin{figure}
\begin{center}
\includegraphics[width=8cm \vspace{-1cm}]{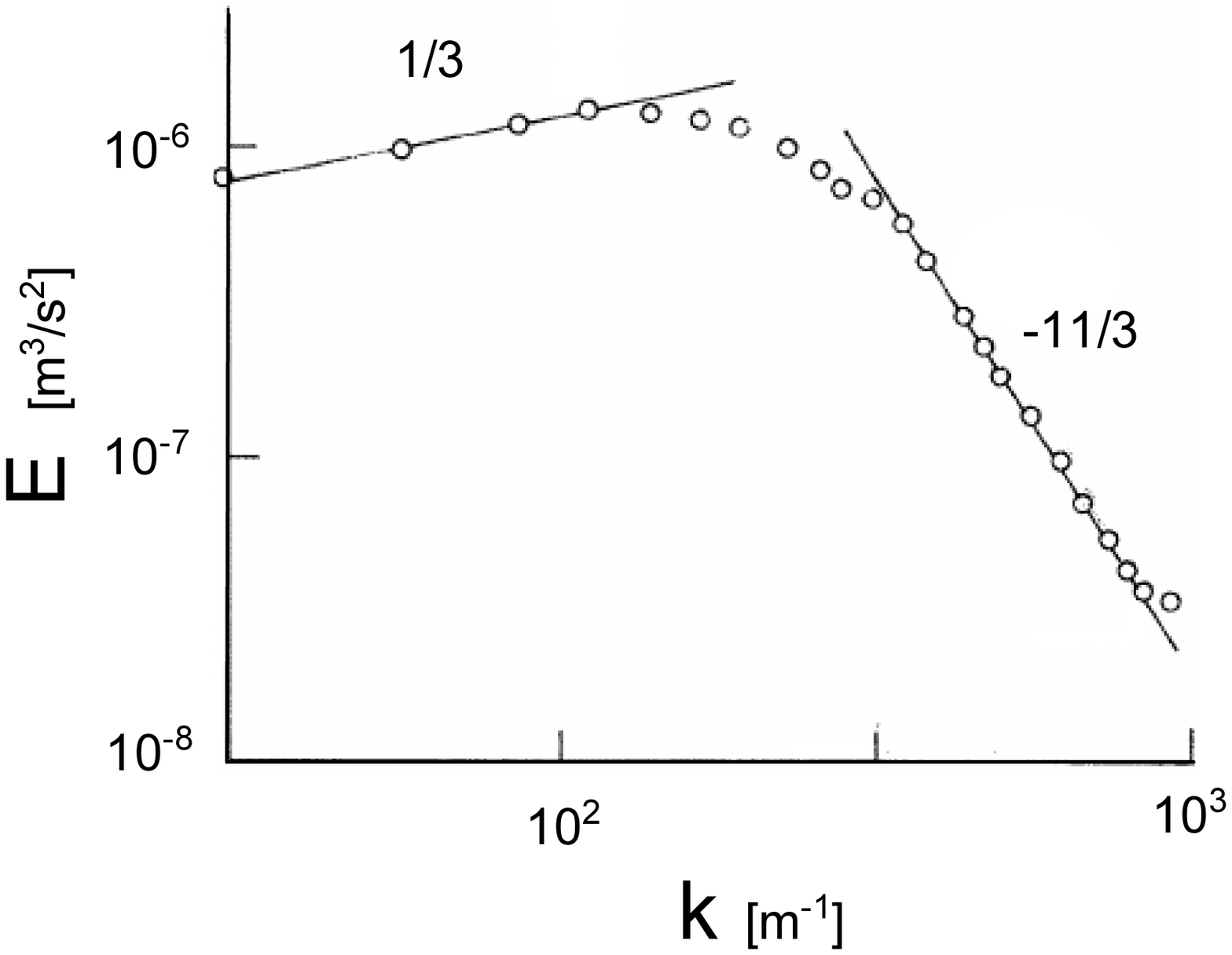}\vspace{-3.4cm}
\caption{\label{fig3} Streamwise power spectrum of the velocity in turbulent flow of mercury past a grid (bars of the grid were inclined in direction of the strong magnetic field). In the log-log scales the straight line with the slope equal to 1/3 corresponds to the scaling Eq. (18)}. 
\end{center}
\end{figure}

\begin{figure}
\begin{center}
\includegraphics[width=8cm \vspace{-1.2cm}]{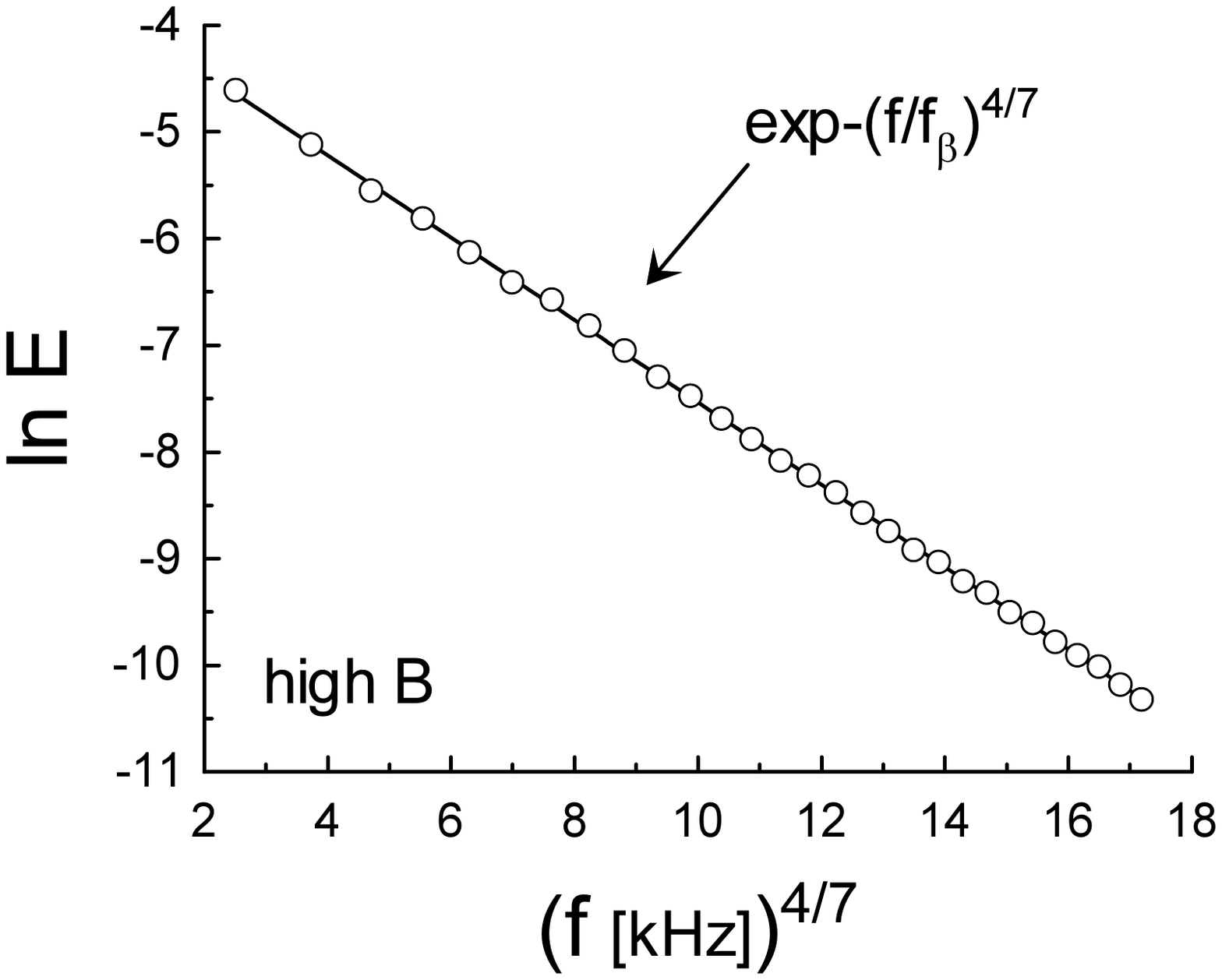}\vspace{-3cm}
\caption{\label{fig4} The same as in Fig. 2 but for high magnetic field.}

\end{center}
\end{figure}

\end{document}